\documentclass[sn-basic, Numbered]{sn-jnl}

\usepackage{graphicx}%
\usepackage{multirow}%
\usepackage{amsmath,amssymb,amsfonts}%
\usepackage{amsthm}%
\usepackage{mathrsfs}%
\usepackage[title]{appendix}%
\usepackage{xcolor}%
\usepackage{textcomp}%
\usepackage{manyfoot}%
\usepackage{booktabs}%
\usepackage{algorithm}%
\usepackage{algorithmicx}%
\usepackage{algpseudocode}%
\usepackage{listings}%

\usepackage{multirow}

\theoremstyle{thmstyleone}%
\theoremstyle{thmstyletwo}%

\theoremstyle{thmstylethree}%

\begin{document}


\title[Article Title]{Nested ResNet: A Vision-Based Method for Detecting the Sensing Area of a Drop-in Gamma Probe}


\author[1]{\fnm{Songyu} \sur{Xu}}

\author[1]{\fnm{Yicheng} \sur{Hu}}
\author[2]{\fnm{Jionglong} \sur{Su}}
\author[1]{\fnm{Daniel} \sur{Elson}}
\author[1,3]{\fnm{Baoru} \sur{Huang}}


\affil[1]{\orgdiv{The Hamlyn Centre for Robotic Surgery}, \orgname{Imperial College London}, \country{UK}}
\affil[2]{\orgdiv{School of AI and Advanced Computing}, \orgname{XJTLU}, \country{China}}
\affil[3]{\orgdiv{Department of Computer Science}, \orgname{University of Liverpool}, \country{UK}}



\abstract{\textbf{Purpose:} Drop-in gamma probes are widely used in robotic-assisted minimally invasive surgery (RAMIS) for lymph node detection. However, these devices only provide audio feedback on signal intensity, lacking the visual feedback necessary for precise localisation. Previous work attempted to predict the sensing area location using laparoscopic images, but the prediction accuracy was unsatisfactory. Improvements are needed in the deep learning-based regression approach. 

\textbf{Methods:} We introduce a three-branch deep learning framework to predict the sensing area of the probe. Specifically, we utilise the stereo laparoscopic images as input for the main branch and develop a Nested ResNet architecture. The framework also incorporates depth estimation via transfer learning and orientation guidance through probe axis sampling. The combined features from each branch enhanced the accuracy of the prediction.

\textbf{Results:} Our approach has been evaluated on a publicly available dataset, demonstrating superior performance over previous methods. In particular, our method resulted in a 22.10\% decrease in 2D mean error and a 41.67\% reduction in 3D mean error. Additionally, qualitative comparisons further demonstrated the improved precision of our approach.

\textbf{Conclusion:} With extensive evaluation, our solution significantly enhances the accuracy and reliability of sensing area predictions. This advancement enables visual feedback during the use of the drop-in gamma probe in surgery, providing surgeons with more accurate and reliable localisation.}




\maketitle

\section{Introduction}\label{sec:introduction}
Cancer \cite{bib1} is a leading cause of death worldwide. Despite this, many cancers offer a high chance of recovery if diagnosed and treated early. Robotic-assisted minimally invasive surgery (RAMIS) is increasingly pivotal in oncological procedures. In comparison to traditional surgical approaches, RAMIS affords greater precision in lesion removal and sparing healthy tissue. In RAMIS surgery, precise targeting of the cancerous tissue is crucial. Leveraging image guidance, surgeons can locate and remove cancerous tissue more precisely and manipulate the surgical robot more efficiently. 

Imaging techniques such as CT, MRI, and PET/SPECT have significantly enhanced preoperative tumour localisation by providing deeper insights \cite{khiewvan2017update}. However, intraoperative manipulations often compromise the accuracy of preoperative imaging due to tissue deformation, presenting challenges in lesion localisation. Consequently, laparoscopic radio-guided tumour detection techniques are particularly critical for intraoperative imaging. This technology employs cancer-target radioactive agents that can emit signals, enabling intraoperative localisation with a gamma probe.

\begin{figure}[htbp]
    \centering
    \includegraphics[width=0.8\linewidth]{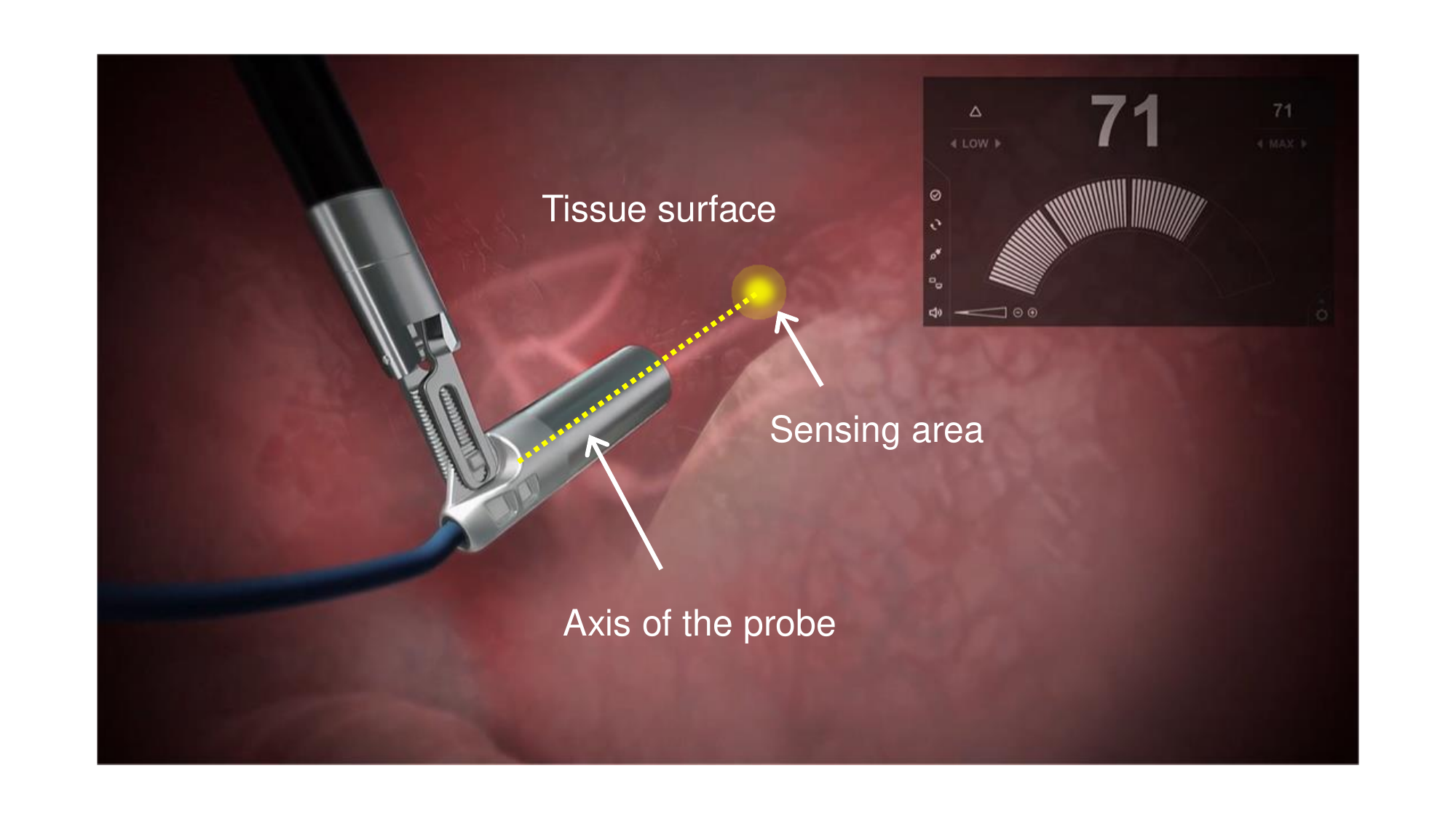}
    \caption{The SENSEI\textsuperscript{\textregistered} gamma probe working scenario and the definition of the sensing area}
    \label{fig:sensing area}
\end{figure}

In RAMIS cancer treatment, drop-in gamma probes are employed \cite{van2024steerable}. SENSEI\textsuperscript{\textregistered} (Lightpoint Medical Ltd) \cite{lightpoint2023sensei} is a typical drop-in gamma probe. This compact device is designed to be compatible with both standard nonarticulated (e.g., laparoscopic Johan grasper) and articulated (e.g., da Vinci ProGrasp) tissue graspers \cite{junquera2022drop} that can be manipulated by surgical robots. The gamma probe can produce auditory feedback according to the signal intensity detected through a small window on its tip. As the intensity increases, the console will produce faster beeps.

Although the audio feedback given by the gamma probe reflects the signal intensity, it cannot indicate the location of the sensing area. To accurately locate the lymph nodes, the surgeon typically needs to ensure the tip of the gamma probe is in contact with the tissue to identify the signal source. However, it may be more convenient to scan by translating or rotating the probe in free space, particularly when operating on a large tissue surface. In addition, during surgical procedures, such as dissection, the gamma probe often loses contact with the tissue surface. As a result, the surgeon must rely on memory to recall the areas previously sensed by the probe and proceed without real-time positional guidance. This can lead to discrepancies between the estimated and actual locations of the sensing area. Hence, the surgeon may need to perform multiple scans to locate the lymph nodes, which increases both the workload and the duration of the procedure. Due to the above limitations, there is a need for an intuitive visual feedback mechanism that can accurately inform the sensing area of the gamma probe when it is not in contact with the tissue. 

Geometrically, the sensing area is defined as the intersection of the extension line of the probe head and tissue surface (Fig. \ref{fig:sensing area}). Estimating the sensing area of the gamma probe can potentially be achieved in three-dimensional (3D) space by calculating the intersection point using geometric information. This approach requires 3D tissue reconstruction and precise information of the probe’s pose. However, obtaining accurate ground truth data for the 3D geometry on both the probe and tissue surface is challenging within the RAMIS context. To address this issue, Huang et al. \cite{huang2023detecting} simplified the complex 3D geometric problem by reducing it to a two-dimensional (2D) one, using a regression model to predict the sensing area location displayed from laparoscopic images. Although their approach was innovative, the basic regression network they utilised demonstrated limited accuracy. Therefore, it is crucial to enhance the prediction model to improve accuracy before advancing to clinical trials.

To improve the accuracy of sensing area location predictions, we have developed a Nested ResNet architecture that incorporates feature guidance. Our approach has been evaluated using the Coffbea dataset \cite{huang2023detecting}, demonstrating superior prediction accuracy compared to previous solutions. Specifically, our contributions are as follows:

\begin{itemize}
    \item We propose a three-branch deep learning framework for predicting the sensing area of the drop-in gamma probe. Relying on stereo laparoscopic images, we introduce a Nested ResNet network for extracting image features that are suitable for the regression task without significant pattern characteristics. We also incorporate feature guidance from depth information and probe axis orientation.

    \item Through extensive experiments, we validate the effectiveness of our method in predicting gamma probe sensing area locations, achieving improvements in accuracy over previous methods.
\end{itemize}

\section{Methodology}\label{sec:methods}
To equip the surgical team with visual guidance displayed on a screen when using the drop-in gamma probe, we introduce an improved method for detecting the sensing area. Relying on stereo laparoscopic images, we propose a Nested ResNet architecture. This is enhanced by combining depth features and orientation features in the other two supplementary branches.

\begin{figure}[htbp]
    \centering
    \includegraphics[width=1\linewidth]{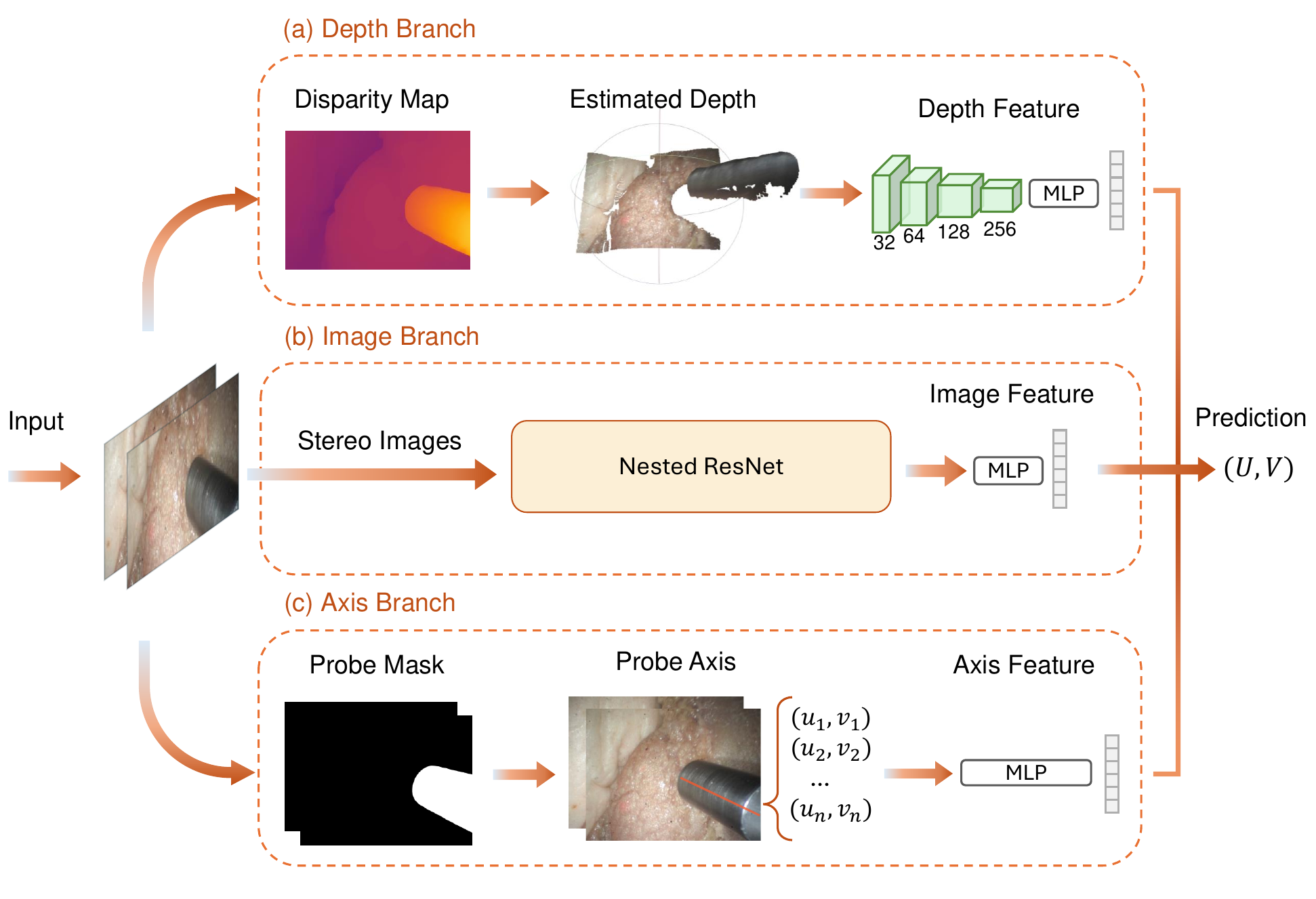}
    \caption{The architecture for sensing area detection: stereo laparoscopic images are input on the left; the main branch (b) uses a Nested ResNet for extracting features from the RGB images; branch (a) shows the depth feature extraction from depth maps using CNN layers; branch (c) shows the feature extraction from axis points using MLP. Features from these three branches are concatenated for predicting the sensing location.}
    \label{fig:network architecture}
\end{figure}

The framework proposed in our work is illustrated in Fig. \ref{fig:network architecture}. The network takes laparoscopic images as input, deriving depth maps and points along the probe axis as supplementary information. Within the image branch, a Nested ResNet structure is utilised, while a convolutional neural network (CNN) processes the depth map inputs, and a multi-layer perceptron (MLP) is employed in the probe axis branch for feature extraction. After extracting features in each of the three branches, features are concatenated to generate the final predictions using MLP. The rest of the section will introduce these three branches in detail.

\subsection{Depth Estimation through Transfer Learning}
Although our solution relies on 2D representation and does not require the explicit construction of 3D models, it is important to understand the spatial relationship between the gamma probe and the background tissue. Therefore we consider incorporating depth maps as an enhancement. Depth estimation can be performed using a pair of stereo images, typically by feature matching on a pair of stereo images \cite{huang2022self}. However, laparoscopic images often have a limited field of view and lack sufficient texture, making stereo matching challenging \cite{huang2021self, huang2022simultaneous}. Alternatively, depth estimation can be achieved using multiple views of a scene \cite{wang2021neus, rey2022360monodepth, wu2022toward, chen2024surgicalgs}, but this requires the surgeon to repeatedly move the laparoscope to obtain adequate views, which reduces the feasibility of this approach during surgical procedures.

Given these challenges, we aim to predict the depth map from a pair of stereo images while addressing the limitations of traditional texture-based stereo matching. To overcome these difficulties, we employ a deep learning-based transfer learning approach for stereo matching and then derive depth maps. Specifically, we utilised the disparity estimation model \cite{xu2023unifying} that can predict the disparity map from a pair of stereo images. This method accepts stereo image pairs as the input and estimates the disparity map using a deep learning approach. It employs convolutional neural networks (CNN) to extract features from each image independently and utilises a transformer-based model to learn feature matching on corresponding images. Since it relies on correspondence matching by comparing the feature similarity of two images, it is not task-dependent and is suitable for transfer learning.

Fig. \ref{fig:network architecture} (a) shows the integration of depth maps in our model. We utilise the disparity estimation model that has been pretrained on three stereo datasets for predicting the disparity map: KITTI Stereo \cite{menze2015object}; Middlebury \cite{scharstein2014high}; and ETH3D Stereo \cite{schops2017multi}. After obtaining the disparity $D$, the per-pixel depth $Z$ can be calculated using $Z = fb/D$, where $f$ is the camera's focal length and $b$ is the baseline. The depth map is fed into a CNN network with MLP for feature extraction and concatenated with features from the other branches.

\subsection{Orientation Guidance Using Central Axis}
As illustrated in Fig. \ref{fig:sensing area}, the sensing area is the intersection of the probe axis and the tissue surface. The probe axis contains key information of the orientation of the sensing area, hence we address the orientation information by introducing the extracted probe axis as supplementary input to our network (Fig. \ref{fig:network architecture} (c)). The central axis of the probe is determined by exploiting its symmetry. The mask of the probe is first generated from the RGB image using Segment Anything Model (SAM) \cite{kirillov2023segany}. Principal Component Analysis (PCA) can then determine the central axis as the first principal component since the greatest variance occurs near the axis of symmetry \cite{huang2023detecting}. A random sample of 50 points along this axis was used to represent the probe orientation and a MLP is applied to extract axis features. These were then concatenated with features from the other two branches.

\subsection{Nested ResNet}
\label{sec:nested resnet}
While the depth branch (Fig. \ref{fig:network architecture} (a)) and the axis branch (Fig. \ref{fig:network architecture} (c)) provides key features for understanding the spatial relationship, it is necessary to extract features directly from original RGB images as they contain all information (Fig. \ref{fig:network architecture} (b)). One challenge of the intersection estimation task is that the sensing area does not have any pattern characteristics but is jointly defined by the relationship of the probe and the background. To address this, we developed a Nested ResNet architecture for the image branch to extract features from the stereo laparoscopic images. This architecture leverages the strengths of the original ResNet model \cite{he2016deep}, incorporating deep residual learning for efficient feature representation. Additionally, we integrate global information using global skip connections, inspired by the UNet \cite{ronneberger2015u}. This design allows for the consideration of both local and global contexts, thereby enhancing the accuracy and robustness of the feature extraction process.

\begin{figure}
    \centering
    \includegraphics[width=1\linewidth]{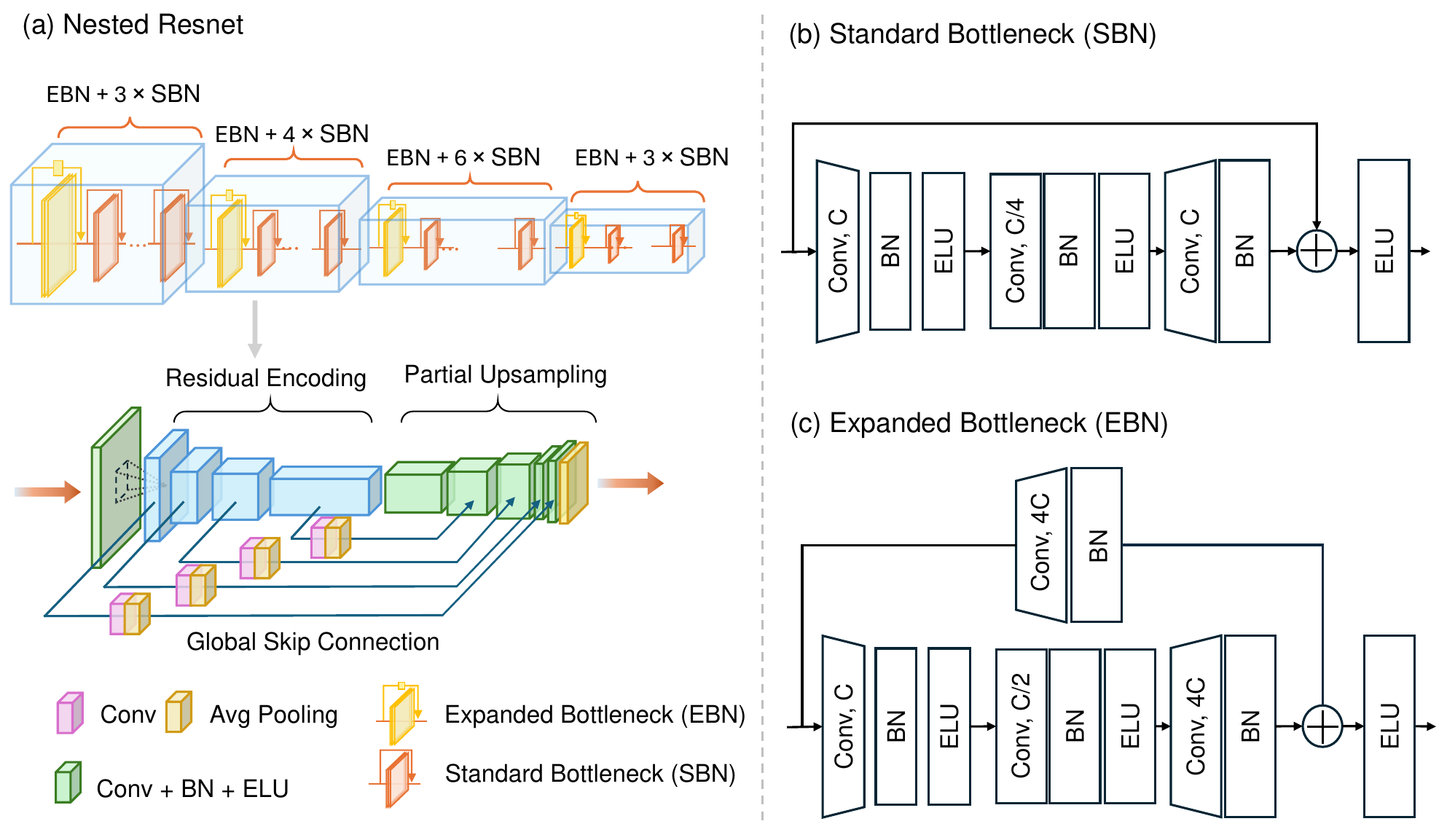}
    \caption{The Nested ResNet design: (a) The overall architecture. (b-c) The standard bottleneck block (SBN) and the expanded bottleneck block (EBN) that form the residual encoding part in the Nested ResNet. The SBN keeps the same channel number $C$ while the EBN expands the channel number from $C$ to $4C$.}
    \label{fig:nested resnet}
\end{figure}

\subsubsection{Residual Encoding Modules}
\label{sec:residual encoding modules}
left and right images are concatenated, resulting in a 6-channel ($C=6$) image of height $H$ and width $W$. To process the image data, a 2D convolutional layer with max pooling is applied. This reduces the spatial dimensions of the image to half, and expands the number of channels to $C=64$. This processed image is then directed into four residual encoding modules as illustrated in Fig. \ref{fig:nested resnet} (a). Each residual encoding module is comprised of two types of residual blocks: the standard bottleneck block and the expanded bottleneck block.

The standard bottleneck residual block structure is widely used in classic networks such as ResNet-50, ResNet-101, and ResNet-152 \cite{he2016deep}. In a standard bottleneck residual block (Fig. \ref{fig:nested resnet} (b)) we apply a $1 \times 1$ convolutional block for reducing the number of channels to $C/4$ for fast computation. After batch normalisation and ELU activation, the middle $3 \times 3$ convolutional block is followed. Finally, a $1 \times 1$ convolutional layer recovers the channel number to the input size and an identity mapping is applied.

In addition to the standard bottleneck residual block, we introduce a variation known as the expanded bottleneck block. Unlike the standard block, which preserves the channel count between its input and output to facilitate identity mapping, the expanded bottleneck block increases the channel depth to augment the network's capacity for feature representation. The compression of these expanded channels for feature extraction is deferred to the decoder part of the network where global skip connections are applied. As shown in Fig. \ref{fig:nested resnet} (c), the expanded bottleneck block initially reduces the channel number to $C/2$ in the middle convolutional layer, but then increases it to $4C$ in the final $1 \times 1$ convolutional block. Replacing the identity mapping, a $1 \times 1$ convolutional block is added to the skip connection to directly expand the channel number of the input from $C$ to $4C$.

Utilising the two types of residual blocks described above, we can construct downsampling modules in the Nested ResNet. Each module includes one expanded bottleneck block to facilitate downsampling and channel expansion, coupled with multiple standard bottleneck blocks designed to deepen the layers, enhancing feature extraction capabilities. In total, our design comprises four downsampling modules, each configured with a different number of expanded bottleneck blocks: 3, 4, 6, and 3, respectively, as illustrated in Fig. \ref{fig:nested resnet} (a).

\subsubsection{Global Skip Connections}
In addition to the residual skip connections utilised in the encoder, additional skip connections bridge the encoder and decoder segments in the Nested ResNet (Fig. \ref{fig:nested resnet}). The sensing area detection, which is crucially influenced by both the probe and the background tissues, requires not just detailed information but also global information. To address this, we have implemented global skip connections. These are designed to preserve and transmit high-resolution details from earlier layers to later stages of the network, ensuring that critical textural and structural information is maintained.

The design of global skip connections is inspired by U-Net architecture \cite{ronneberger2015u}. Within these connections, a $1 \times 1$ convolution is used to adjust the channel number, while adaptive average pooling modifies the spatial dimensions to suit the decoder's requirements. As for the decoder, after the residual encoding modules, partial upsampling is applied. In contrast to traditional U-Net implementations, which recover the spatial dimension to the original size for whole image segmentation, our approach uses partial upsampling for the regression task for the regression task. Each upsampling block only doubles the spatial size to save computing expenses. This configuration allows high-resolution details to bypass multiple intermediate downsampling stages, using only two convolutional layers in the global skip connections applied to preserve more comprehensive information.

\subsection{Loss Function}
We use the square of the Euclidean distance between the ground truth and prediction as the loss function:
\begin{equation}
    \mathcal{L} = (p_x - q_x)^2 + (p_y - q_y)^2
\end{equation}
where $(p_x, p_y)$ is the predicted location and $(q_x, q_y)$ is the ground truth value.

\section{Experiments and Results}

\subsection{Dataset}
\begin{figure}[htbp]
    \centering
    \includegraphics[width=1\linewidth]{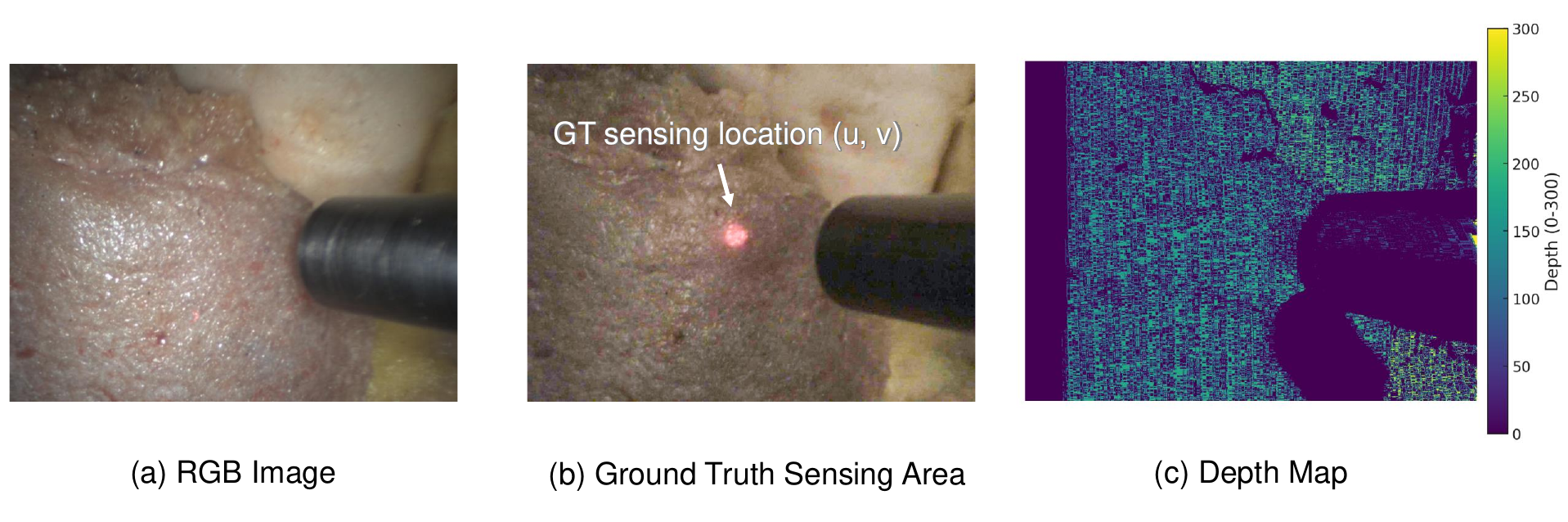}
    \caption{The Coffbea \cite{huang2023detecting} dataset that contains three types of data: (a) RGB laparoscopic image, (b) ground truth sensing area provided in 2D coordinates, and (c) pre-generated depth maps.}
    \label{fig:dataset and features}
\end{figure}

The Coffbea dataset \cite{huang2023detecting}, as illustrated in Fig. \ref{fig:dataset and features}, comprises images capturing various operational scenarios of the SENSEI\textsuperscript{\textregistered} gamma probe. The probe is held by a ProGrasp forcep and images were taken using a stereo laparoscopic camera. Images of various perspectives of the gamma probe during simulated operation on a silicone model were captured. The fraction of the probe appearing in the image varies from 50\%-100\%. The ground truth location for the sensing area was established using a fabricated probe with a laser module embedded in its casing. Therefore, the point illuminated by the laser confirms where the tissue surface intersects with the central axis of the probe and was marked as the ground truth sensing area (Fig. \ref{fig:dataset and features} (b)). The dataset also includes pre-generated depth maps for each image, which facilitate evaluation (Fig. \ref{fig:dataset and features} (c)).

\subsection{Experimental Settings}
The Coffbea dataset \cite{huang2023detecting} was used for experiments, with the samples divided into 800 training images, 200 validation images, and 200 test images. Each image was $920\times1224$ in size and was padded to $1224 \times 1224$. Our solution was implemented with PyTorch on an Ubuntu operating system. Experiments were conducted on an NVIDIA A40 GPU with 48 GB memory, a batch size of 8, and an Adam optimiser \cite{kingma2014adam} with an initial learning rate of $1 \times 10^{-4}$, decaying to $8 \times 10^{-5}$. We trained the network for 300 epochs.

\subsection{Evaluation Metrics}
We calculated the Euclidean distances in the 2D space between the predicted sensing area locations and the ground truth values.
\begin{equation}
    L_{2d} = \sqrt{(u_{\text{pred}} - u_{\text{gt}})^2 + (v_{\text{pred}} - v_{\text{gt}})^2}
\end{equation}
where $(u, v)$ is the location in 2D plane and the distance is measured in pixels. The mean error, standard deviation, and median error were reported. 

Moreover, the corresponding depth information of any 2D point can be extracted from the depth map. Subsequently, both predicted and ground-truth 3D values can be obtained by back-projecting their 2D coordinates to 3D using the depth value and the camera's intrinsic parameters:

\begin{equation}
X = \frac{(u - o_x) Z}{\alpha},\quad
Y = \frac{(v - o_y) Z}{\beta},\quad
Z = Z
\end{equation}
where $(u, v)$ is the location in the 2D coordinate system, and $Z$ is the depth value acquired from the depth map; $\alpha$ and $\beta$ are the scaling factors in the $X$ and $Y$ directions; $(o_x, o_y)$ are the coordinates of the camera's optical centre. Hence, we can calculate the 3D distance:
\begin{equation}
    L_{3d} = \sqrt{(X_{\text{pred}} - X_{\text{gt}})^2 + (Y_{\text{pred}} - Y_{\text{gt}})^2 + (Z_{\text{pred}} - Z_{\text{gt}})^2}
\end{equation}

\subsection{Quantitative Experiments}
Table \ref{tab:quantitative table} gives the quantitative evaluation of our methods with previous solution SL Regress from Huang et al. \cite{huang2023detecting}. In our solution, we utilise laparoscopic images as the primary input, from which we derive two key features: points along the probe axis and an estimated depth map. Notably, depth estimation is a novel addition not included in previous research. Furthermore, our solution is implemented using the Nested ResNet (NResNet) architecture, which we designed specifically for the sensing area prediction task, whereas the baseline method SL Regress \cite{huang2023detecting} only combines traditional networks such as ResNet-50 \cite{he2016deep}, Vision Transformer (ViT) \cite{dosovitskiy2020image}, multi-layer perceptron (MLP) and long short-term memory (LSTM) \cite{hochreiter1997long}. Our method, employing Nested ResNet as the backbone and incorporating probe axis and depth estimation, outperforms all others, achieving the lowest 2D mean error with a reduction of 22.10\% compared to the best results from previous studies. The corresponding 3D mean error is also the lowest, achieving a reduction of 41.67\%. However, using only the probe axis as supplementary information does not necessarily lead to improved outcomes. This may be attributed to the fact that the probe axis primarily provides linear orientation information without spatial context, potentially leading to the learning of noisy features. In contrast, the estimated depth map offers crucial spatial information, facilitating the detection of the intersection between the probe axis and the surface.

\begin{table}[htbp]
\caption{Quantitative comparisons of our solution with baseline methods, SL Regress \cite{huang2023detecting}. The best results achieved by our methods are shown in \textbf{bold}, while the best results from the previous method SL Regress are \underline{underlined}. 2D and 3D errors are measured in pixels and millimetres respectively}
\label{tab:quantitative table}
\centering
\renewcommand{\arraystretch}{1.25}
\begin{tabular}{c|ccc|ccc|ccc}
\toprule
& \multicolumn{3}{c|}{Method} & \multicolumn{3}{c|}{2D Metrics} & \multicolumn{3}{c}{3D Metrics} \\ \hline
 &Image     &  Axis & Depth & Mean E. &   STD &  Median &   Mean E. &  STD  &  Median\\ \hline
\multirow{6}{*}{\rotatebox{90}{SL Regress \cite{huang2023detecting}}}&ResNet-50 &   -   &   -    &   85.0  &  65.0  &  66.9  & 10.7  &  15.9  &  6.1 \\
                  &ResNet-50 & LSTM  &   -    &  81.0  &   63.4 &  67.8  &  10.6 &  15.6 &  7.2 \\
                  &ResNet-50 &  MLP  &   -    &  \underline{55.2}   &  \underline{44.7} &  \underline{40.9}  &  12.7 &  21.7 &  6.6 \\
                  &ViT       &   -   &   -    &  66.3   & 65.4  &  49.1  &  8.9 &   15.6 &  4.7 \\
                  &ViT       & LSTM  &   -    &  73.8   & 68.3  &  57.0  &  \underline{6.0}  &  \underline{6.1}  &  \underline{3.7} \\
                  &ViT       &  MLP  &   -    &  92.2   & 73.1  &  72.1  &  11.3  & 21.8  &  5.6  \\ \hline
\multirow{3}{*}{\rotatebox{90}{Ours}} &NResNet   &   -   &    -  &  47.8    &  32.5  &  40.4  &  7.6  &  14.0 &  4.1 \\
                  &NResNet   &  MLP  &   -    &   51.5  &  34.4  &  40.0   &  8.7  &  16.6  &  3.2  \\
                  &NResNet   &  -  &   CNN    &   44.4   & 30.6  &  39.6  &  9.2  &  17.0  &  3.7 \\
                  &NResNet   &  MLP  &   CNN    & \textbf{43.0}    & \textbf{30.5}  & \textbf{36.9}   & \textbf{3.5}  &  \textbf{2.6}  &  \textbf{2.7} \\ \bottomrule
\end{tabular}
\end{table}

\subsection{Qualitative Evaluation}
\begin{figure}[h]
    \centering
    \includegraphics[width=1\linewidth]{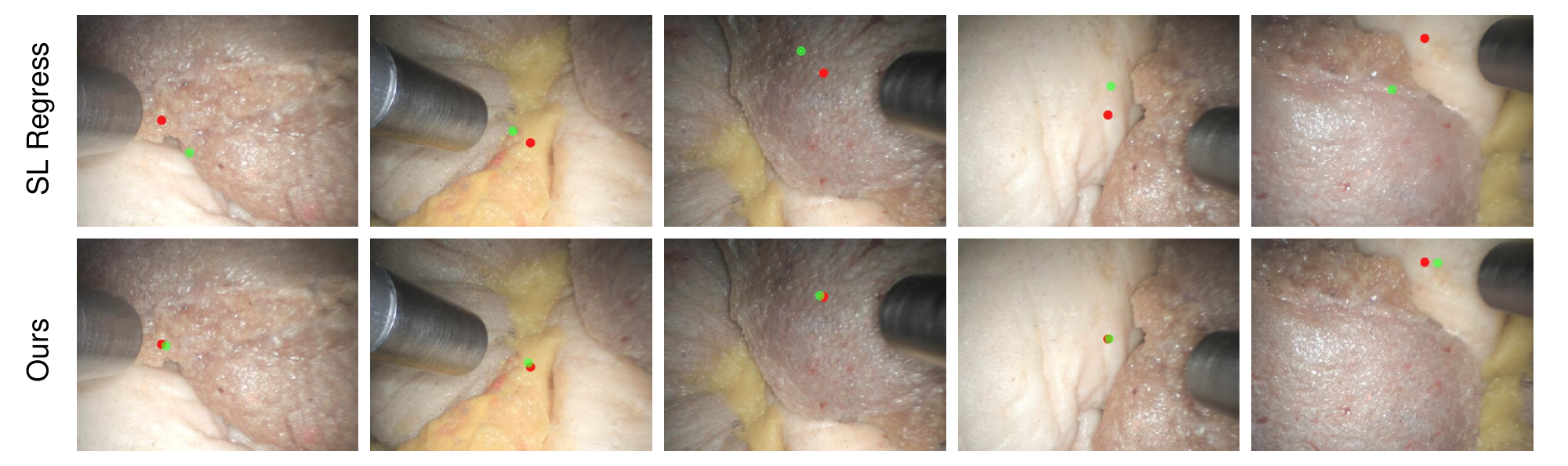}
    \caption{Visualisation. Red dots indicate the location of the ground truth and green dots stand on the location of the prediction. The top row shows the results using previous method SL Regress \cite{huang2023detecting}, and the bottom row shows the results of our methods}
    \label{fig:visualisation}
\end{figure}

Visualisation examples of the sensing point prediction can be seen in figure \ref{fig:visualisation}. We mark the predicted location on the image with green dots while the ground truth location is also marked in red for comparison. The top row gives the best results from previous work \cite{huang2023detecting} and the bottom row shows results using our method. It can be seen from the qualitative comparisons that our predictions achieve higher accuracy.

\section{Conclusion}
In this study, we aimed to enhance the prediction accuracy of the sensing area for a non-imaging drop-in gamma probe. We introduced a three-branch deep learning framework for this regression task, utilising stereo laparoscopic images as input and generating depth maps and the axes of the probe as additional features to provide more detailed geometric information. In addition, we designed a Nested ResNet architecture for extracting task-related features from the RGB images. This architecture benefited from both the residual downsampling modules for detailed feature extraction and the global skip connection for retaining high-resolution information. Extensive experiments demonstrated the effectiveness of our approach. In conclusion, our work marks a significant improvement in predicting the sensing area of drop-in gamma probes, refining accuracy beyond previous benchmarks. This advancement will be crucial for accurate lymph node identification and tumour localisation in minimally invasive cancer treatment, enhancing the effectiveness of intraoperative procedures.






\bmhead{Acknowledgements}
This work was supported in part by the U.K. National Institute for Health Research (NIHR) Invention for Innovation under Award NIHR200035. It is independent research funded by the National Institute for Health Research (NIHR) Imperial Biomedical Research Centre (BRC), and the Cancer Research UK (CRUK) Imperial Centre.

\bibliography{sn-bibliography}

\begin{thebibliography}{23}
\providecommand{\natexlab}[1]{#1}
\providecommand{\url}[1]{{#1}}
\providecommand{\urlprefix}{URL }
\providecommand{\doi}[1]{\url{https://doi.org/#1}}
\providecommand{\eprint}[2][]{\url{#2}}
 \bibcommenthead

\bibitem[{Campbell and Gear(1995)}]{bib1}
Campbell SL, Gear CW (1995) The index of general nonlinear {D}{A}{E}{S}. Numer {M}ath 72(2):173--196

\bibitem[{Chen et~al(2024)Chen, Zhang, Islam, Vasconcelos, Stoyanov, Elson, and Huang}]{chen2024surgicalgs}
Chen J, Zhang X, Islam M, et~al (2024) Surgicalgs: Dynamic 3d gaussian splatting for accurate robotic-assisted surgical scene reconstruction. arXiv preprint arXiv:241009292

\bibitem[{DOSOVITSKIY(2020)}]{dosovitskiy2020image}
DOSOVITSKIY A (2020) An image is worth 16x16 words: Transformers for image recognition at scale. arXiv preprint arXiv:201011929

\bibitem[{He et~al(2016)He, Zhang, Ren, and Sun}]{he2016deep}
He K, Zhang X, Ren S, et~al (2016) Deep residual learning for image recognition. In: Proceedings of the IEEE conference on computer vision and pattern recognition, pp 770--778

\bibitem[{Hochreiter(1997)}]{hochreiter1997long}
Hochreiter S (1997) Long short-term memory. Neural Computation MIT-Press

\bibitem[{Huang et~al(2021)Huang, Zheng, Nguyen, Tuch, Vyas, Giannarou, and Elson}]{huang2021self}
Huang B, Zheng JQ, Nguyen A, et~al (2021) Self-supervised generative adversarial network for depth estimation in laparoscopic images. In: Medical Image Computing and Computer Assisted Intervention--MICCAI 2021: 24th International Conference, Strasbourg, France, September 27--October 1, 2021, Proceedings, Part IV 24, Springer, pp 227--237

\bibitem[{Huang et~al(2022{\natexlab{a}})Huang, Nguyen, Wang, Wang, Mayer, Tuch, Vyas, Giannarou, and Elson}]{huang2022simultaneous}
Huang B, Nguyen A, Wang S, et~al (2022{\natexlab{a}}) Simultaneous depth estimation and surgical tool segmentation in laparoscopic images. IEEE transactions on medical robotics and bionics 4(2):335--338

\bibitem[{Huang et~al(2022{\natexlab{b}})Huang, Zheng, Nguyen, Xu, Gkouzionis, Vyas, Tuch, Giannarou, and Elson}]{huang2022self}
Huang B, Zheng JQ, Nguyen A, et~al (2022{\natexlab{b}}) Self-supervised depth estimation in laparoscopic image using 3d geometric consistency. In: International Conference on Medical Image Computing and Computer-Assisted Intervention, Springer, pp 13--22

\bibitem[{Huang et~al(2023)Huang, Hu, Nguyen, Giannarou, and Elson}]{huang2023detecting}
Huang B, Hu Y, Nguyen A, et~al (2023) Detecting the sensing area of a laparoscopic probe in minimally invasive cancer surgery. In: International Conference on Medical Image Computing and Computer-Assisted Intervention, Springer, pp 260--270

\bibitem[{Junquera et~al(2022)Junquera, Harke, Walz, Hadaschik, Adshead, Everaerts, Goffin, Grootendorst, Oldfield, Vyas et~al}]{junquera2022drop}
Junquera JMA, Harke NN, Walz JC, et~al (2022) A drop-in gamma probe for minimally invasive sentinel lymph node dissection in prostate cancer: preclinical evaluation and interim results from a multicenter clinical trial. Clinical Nuclear Medicine pp 10--1097

\bibitem[{Khiewvan et~al(2017)Khiewvan, Torigian, Emamzadehfard, Paydary, Salavati, Houshmand, Werner, and Alavi}]{khiewvan2017update}
Khiewvan B, Torigian DA, Emamzadehfard S, et~al (2017) An update on the role of pet/ct and pet/mri in ovarian cancer. European journal of nuclear medicine and molecular imaging 44:1079--1091

\bibitem[{Kingma(2014)}]{kingma2014adam}
Kingma DP (2014) Adam: A method for stochastic optimization. arXiv preprint arXiv:14126980

\bibitem[{Kirillov et~al(2023)Kirillov, Mintun, Ravi, Mao, Rolland, Gustafson, Xiao, Whitehead, Berg, Lo et~al}]{kirillov2023segany}
Kirillov A, Mintun E, Ravi N, et~al (2023) Segment anything. In: Proceedings of the IEEE/CVF International Conference on Computer Vision, pp 4015--4026

\bibitem[{Medical(2021)}]{lightpoint2023sensei}
Medical L (2021) Miniature surgical gamma probe. \urlprefix\url{https://senseisurgical.com/}

\bibitem[{Menze and Geiger(2015)}]{menze2015object}
Menze M, Geiger A (2015) Object scene flow for autonomous vehicles. In: Proceedings of the IEEE conference on computer vision and pattern recognition, pp 3061--3070

\bibitem[{van Oosterom et~al(2024)van Oosterom, Diaz-Feij{\'o}o, Santisteban, S{\'a}nchez-Izquierdo, Perissinotti, Glickman, Marina, Torn{\'e}, van Leeuwen, and Vidal-Sicart}]{van2024steerable}
van Oosterom MN, Diaz-Feij{\'o}o B, Santisteban MI, et~al (2024) Steerable drop-in radioguidance during minimal-invasive non-robotic cervical and endometrial sentinel lymph node surgery. European Journal of Nuclear Medicine and Molecular Imaging pp 1--9

\bibitem[{Rey-Area et~al(2022)Rey-Area, Yuan, and Richardt}]{rey2022360monodepth}
Rey-Area M, Yuan M, Richardt C (2022) 360monodepth: High-resolution 360deg monocular depth estimation. In: Proceedings of the IEEE/CVF Conference on Computer Vision and Pattern Recognition, pp 3762--3772

\bibitem[{Ronneberger et~al(2015)Ronneberger, Fischer, and Brox}]{ronneberger2015u}
Ronneberger O, Fischer P, Brox T (2015) U-net: Convolutional networks for biomedical image segmentation. In: Medical image computing and computer-assisted intervention--MICCAI 2015: 18th international conference, Munich, Germany, October 5-9, 2015, proceedings, part III 18, Springer, pp 234--241

\bibitem[{Scharstein et~al(2014)Scharstein, Hirschm{\"u}ller, Kitajima, Krathwohl, Ne{\v{s}}i{\'c}, Wang, and Westling}]{scharstein2014high}
Scharstein D, Hirschm{\"u}ller H, Kitajima Y, et~al (2014) High-resolution stereo datasets with subpixel-accurate ground truth. In: Pattern Recognition: 36th German Conference, GCPR 2014, M{\"u}nster, Germany, September 2-5, 2014, Proceedings 36, Springer, pp 31--42

\bibitem[{Schops et~al(2017)Schops, Schonberger, Galliani, Sattler, Schindler, Pollefeys, and Geiger}]{schops2017multi}
Schops T, Schonberger JL, Galliani S, et~al (2017) A multi-view stereo benchmark with high-resolution images and multi-camera videos. In: Proceedings of the IEEE conference on computer vision and pattern recognition, pp 3260--3269

\bibitem[{Wang et~al(2021)Wang, Liu, Liu, Theobalt, Komura, and Wang}]{wang2021neus}
Wang P, Liu L, Liu Y, et~al (2021) Neus: Learning neural implicit surfaces by volume rendering for multi-view reconstruction. arXiv preprint arXiv:210610689

\bibitem[{Wu et~al(2022)Wu, Wang, Hall, Neumann, and Su}]{wu2022toward}
Wu CY, Wang J, Hall M, et~al (2022) Toward practical monocular indoor depth estimation. In: Proceedings of the IEEE/CVF conference on computer vision and pattern recognition, pp 3814--3824

\bibitem[{Xu et~al(2023)Xu, Zhang, Cai, Rezatofighi, Yu, Tao, and Geiger}]{xu2023unifying}
Xu H, Zhang J, Cai J, et~al (2023) Unifying flow, stereo and depth estimation. IEEE Transactions on Pattern Analysis and Machine Intelligence

\end{thebibliography}

\end{document}